\documentclass[aps,prl,showpacs,twocolumn,amsmath,amssymb,nofootinbib]{revtex4-1} 
\usepackage{graphicx}
\usepackage{placeins}

\usepackage{dcolumn}
\usepackage{bm}
\usepackage{epsf}

\newcommand{\aap}{\textnormal{A\&A}}
\newcommand{\mnras}{\textnormal{MNRAS}}
\newcommand{\apjs}{\textnormal{ApJS}}
\newcommand{\apjl}{\textnormal{ApJL}}
\newcommand{\na}{New Astronomy}

\begin{document}
\title{Correlating Fourier phase information with real-space higher order statistics}
\date{\today}

\author{H. I. Modest$^1$\email{hmodest@mpe.mpg.de}, C. R\"ath$^1$, A. J. Banday$^{2,3}$, K. M. G\'{o}rski$^{4,5}$ and  G. E. Morfill$^1$}
\affiliation{$^1$ Max-Planck Institut f\"ur extraterrestrische Physik, PO Box 1312, Giessenbachstr. 1, D-85748 Garching, Germany\\
        $^2$ CNRS, IRAP, 9 Av. Colonel Roche, BP 44346, F-31028 Toulouse CEDEX 4, France\\
        $^3$ Universit\'e de Toulouse, UPS-OMP, IRAP,  F-31028 Toulouse CEDEX 4, France\\
        $^4$Jet Propulsion Laboratory, California Institute of Technology, Pasadena, CA 91109, USA\\
        $^5$Warsaw University Observatory, Aleje Ujazdowskie 4, 00 - 478 Warszawa, Poland}


\begin{abstract}
We establish for the first time heuristic correlations between harmonic space phase information and higher order statistics.
Using the spherical full-sky maps of the cosmic microwave background as an example we demonstrate that known phase correlations at large spatial scales can gradually be diminished 
when subtracting a suitable best-fit (Bianchi-) template map of given strength. The weaker phase correlations lead in turn 
to a vanishing signature of anisotropy when measuring the Minkowski functionals and scaling indices in real-space and 
comparing them with surrogate maps being free of phase correlations. Those investigations can open a new road to a better understanding of signatures of non-Gaussianities in complex spatial structures
by elucidating the meaning of Fourier phase correlations and their influence on higher order statistics.
\end{abstract}
\maketitle
\textit{Introduction}.---Great advances in imaging techniques nowadays allow for a visualization of spatial structures in medicine, biochemistry, solid state physics or astronomy ranging from the atomic nanometer scale \cite[AF Microscopy,][]{PhysRevLett.56.930} to megaparsec for the large scale structure of the Universe \cite[SDSS DR9,][]{2012ApJS..203...21A} with the largest and oldest observable structure being the cosmic microwave background (CMB) \cite{1992ApJ...396L...1S,2003ApJS..148....1B,2013arXiv1303.5062P}. These images of complex natural structures contain a wealth of information about the origin and -- often nonlinear -- formation process of structure. As known from the field of signal processing and imaging science for example in optics, cybernetics or time-series analysis, a comprehensive analysis of signals stemming from systems with nonlinear dynamics must go beyond a linear analysis (autocorrelation function in real-space or the power spectrum in Fourier/harmonic space). In image analysis by higher order statistics (HOS) this is only achieved when the phase information is included. A detailed understanding of this phase information has become very important in natural sciences in recent years and can improve existing methods of image analysis, image reconstruction and also image compression \cite{PhysRevLett.98.034801}. Examples for studies on phase information and their application can be found in the development of the first phase retrieval methods in X-ray imaging in the last century \cite{ISI:A1972M312400012,Fienup:82,0034-4885-54-11-002} or in studies of the phase distribution in CMB data \cite{2004MNRAS.350..989C,2007MNRAS.380L..71C,2007ApJ...664....8C}, inter alia. 

The higher order $n$-point correlation functions with $n > 2$ in real-space or their equivalent polyspectra in Fourier space, however, do not allow direct conclusions on the distribution of the phases yet. If the signal is Gaussian, the Fourier phases are independent and identically uniform distributed. In this case, the second order measures are anyway sufficient to understand the underlying physics. In generic cases though, in which the underlying random fields are non-Gaussian, the phases are correlated and contain information that must not be neglected. One of the next steps on the way to a more profound understanding of images in general is a detailed description of the phase distribution and the investigation of the relation between phase information and real-space HOS.

In cosmology, the search for primordial non-Gaussian random fields has attracted great attention because their detection and identification allows for a differentiation between various models of inflation. While e.g. multi-field inflation or self-interactions of the inflaton field generally yield measurable non-Gaussianity (NG), the standard isotropic cosmology with the simple single-field slow-roll inflationary scenario and a Friedmann-Robertson-Walker (FRW) metric predicts a Gaussian distribution of the first density perturbations of the Universe \cite{1981PhRvD..23..347G,1982PhRvL..48.1220A,1982PhLB..108..389L}. The latest, most precise measurements of parametrized NGs of the local, equilateral and orthogonal type by the Planck team did not reveal significant deviations from Gaussianity \cite{2013arXiv1303.5084P}. A model independent test using the well-established method of surrogates \cite{Theiler199277} applied to the Planck CMB maps revealed, however, NGs or hemispherical asymmetries for higher order statistics \cite{2013arXiv1303.5083P}, which can be traced back to harmonic space phase correlations on large spatial scales at low spherical harmonic modes $\ell$ with $\ell < 20$, confirming previous findings in WMAP data  \cite{2011MNRAS.415.2205R,2012PhRvD..86h3005R,2013MNRAS.428..551M}. Evidence was found that a best-fit Bianchi type $\mathrm{VII_h}$ template (BT) correlates with the large-scale anomalies in the CMB sky  \cite{2005ApJ...629L...1J,2006MNRAS.369..598C,McEwen11072006}, although the best-fit Bianchi model itself is not compatible with the parameters of the cosmological concordance model  (see e.g. \cite{2006ApJ...644..701J}). Bianchi models provide a generic description of anisotropic homogeneous cosmologies  \cite{Barrow:1985vn} that are only asymptotically close to a FRW universe. Applying a BT correction to CMB data yields a sky which is statistically isotropic for at least some subset of statistical measures, e.g. the local power estimates. In \cite{2013arXiv1303.5083P}, it was found that the signal stemming from low-$\ell$ phase correlations can also be significantly reduced if the best-fitting BT is subtracted from the Planck maps. These results could hint at the properties of fully compliant cosmological models, especially when the behavior of the data is studied as a function of the correction and on isolated scales.

This Letter aims at the systematic investigation of the fundamental relation between the Fourier phase distribution in harmonic space and real-space higher order statistics, comparable to the Wiener-Khinchin theorem \cite{Wiener:1930fk,citeulike:2815228}. To do so, we analyze the distributions of Fourier phases directly using a non-parametric statistical test called Kuiper test and compare these results to real-space signatures from higher order correlations involving surrogates maps. The investigations are carried out using the CMB as an example of a spherical data set where no boundary conditions have to be met. We find that our previous real-space results studying the phase correlations of CMB data on the large scales with $\ell$-modes of $\ell < 20$ are reproduced in a study with $\ell < 10$. With this even stricter choice of $\ell$-interval we make sure that our analysis of the phase distribution is comparable to \cite{2007ApJ...664....8C} where they have used the exact same range. To modify the strength of phase correlations contained in the data, we make use of anisotropic Bianchi type $\mathrm{VII_h}$ best-fit templates. We compare full sky CMB maps of the \textit{Wilkinson Microwave Anisotropy Probe (WMAP)} experiment \cite{2013ApJS..208...20B} and the \textit{Planck} mission \cite{2013arXiv1303.5062P}.

\textit{Methods}.---Assuming that an image $I(x,y)$ can be represented in terms of linear superposition of (not necessarily orthogonal) basis functions $\beta_i(x,y)$ by $I(x,y) = \sum\limits_{i} a_i \beta_i(x,y)$, the CMB map with its temperature anisotropies $\Delta T/T(\theta,\phi)$ at angular position $(\theta,\phi)$ can be expanded in orthonormal spherical harmonics $Y_{\ell m}$ as   
\begin{equation}
\centering
\Delta T/T(\theta,\phi) = \sum\limits_{l=0}^{\infty} \sum\limits_{m=-l}^{l}a_{\ell m} Y_{\ell m}(\theta,\phi)
\end{equation}
with the complex spherical harmonic coefficients
\begin{equation*}
a_{\ell m} = \int \mathrm{d} \textbf{n}\ T(\textbf{n}) Y^*_{\ell m}(\textbf{n}),
\end{equation*}
where $\textbf{n}$ is the unit direction vector, $T$ is the CMB temperature anisotropy, $Y^*_{\ell m}$ are the complex conjugates of the spherical harmonics, and $a_{\ell m} = | a_{\ell m}| e^{i\varphi_{\ell m}}$. The set of the spherical harmonics is defined with respect to a particular coordinate system. The phases $\varphi_{\ell m}$ of the harmonic coefficients are not rotational invariant.  For the CMB the usual system is in Galactic coordinates.

Gaussianity of the CMB means a Gaussian distribution of its independent complex spherical harmonic coefficients $a_{\ell m}$. According to theory, the harmonic space phases are then independent and identically distributed (i.i.d.) and follow a uniform distribution in the interval $[-\pi,\pi]$. We test for this null-hypothesis of uncorrelated phases with two complementary methods to enable a comparison between them. Method A directly explores the distribution of the phases in harmonic space and is motivated by the findings in \cite{2007ApJ...664....8C}. Method B is based on a real-space analysis using the method of surrogates. Generating the surrogate maps, we destroy only a single characteristic of the original map, which is a possible correlation of the phases. We gradually diminish the strength of phase correlations by subtracting a Bianchi type $\mathrm{VII_h}$ best-fit template (\cite{2006A&A...460..393J} for WMAP, \cite{2013arXiv1303.3409M,2013arXiv1303.5086P} for Planck) multiplied by a strength factor of $a = 0.1, 0.3, 0.5, 0.7,0.9$ and $1.0$ from the original map to enable a comparison of the response in method A and B as a function of the correction. The maps are used as full-sky maps. 

\noindent \textbf{Method A:} For a precise analysis of the phase distributions of the maps we calculate the $a_{\ell m}$ coefficients of Planck SMICA and SEVEM \cite{planckdata} and WMAP 9 ILC \cite{wmapdata} full-sky maps and test their phases $\varphi_{\ell m}$ for independence. We obtain $54$ phases $\varphi_{\ell m}$ with values between $-\pi$ and $\pi$ depending on the chosen coordinate system after an $a_{\ell m}$-decomposition for $\ell \in [2,10]$, $m>0$. If these $\varphi_{\ell m}$ fulfill the random phase hypothesis, i.e. are i.i.d. and follow a uniform distribution, the phase difference taken between these phases should be uniformly distributed in $[0,2\pi]$. To test this we define subsets of differences with fixed separations  $(\Delta \ell,\Delta m)$ by $\Delta \varphi\left ( \Delta \ell, \Delta m \right)= \varphi_{\ell+\Delta \ell,m+\Delta m}-\varphi_{\ell m}$. The Kuiper statistic (KS) \cite{Kuiper1960,citeulike:1386464} is then used to test for the null-hypothesis by comparing the cumulative distribution function (CDF) of the $\Delta \varphi\left ( \Delta \ell, \Delta m \right)$ with a given uniform CDF. The Kuiper test statistic is $V=D^+ + D^-$, where $D^+$ and $D^-$ represent the absolute sizes of the most positive and most negative difference between the two cumulative distribution functions that are being compared. The $p$-value of an observed value $V$ is given by $p=Q_{KP}\left( \left[\sqrt{N_e} + 0.155 +0.24\/\sqrt{N_e}\right] V \right)$ with respect to the monotonic function $Q_{KP}(\lambda)=2 \sum_{j=1}^{\infty}(4j^2\lambda^2-1)e^{-2j^2\lambda^2}$, where $N_e$ is the effective number of data points. It is interpreted as the probability the measured value $V$ has if the null-hypothesis were true. High $p$-values for a given $(\Delta \ell, \Delta m)$ separation therefore indicate the absence of phase-correlations between mode-pairs separated by $(\Delta \ell, \Delta m)$ whereas low values indicate their existence. Values of $p\le 0.05$ (5\% level) are widely accepted as strong evidence against the null-hypothesis. Our results depend on the chosen coordinate system. However, averaging over rotated systems will show a global trend of the results.
\begin{figure*}
\includegraphics[width=15cm,keepaspectratio=true]{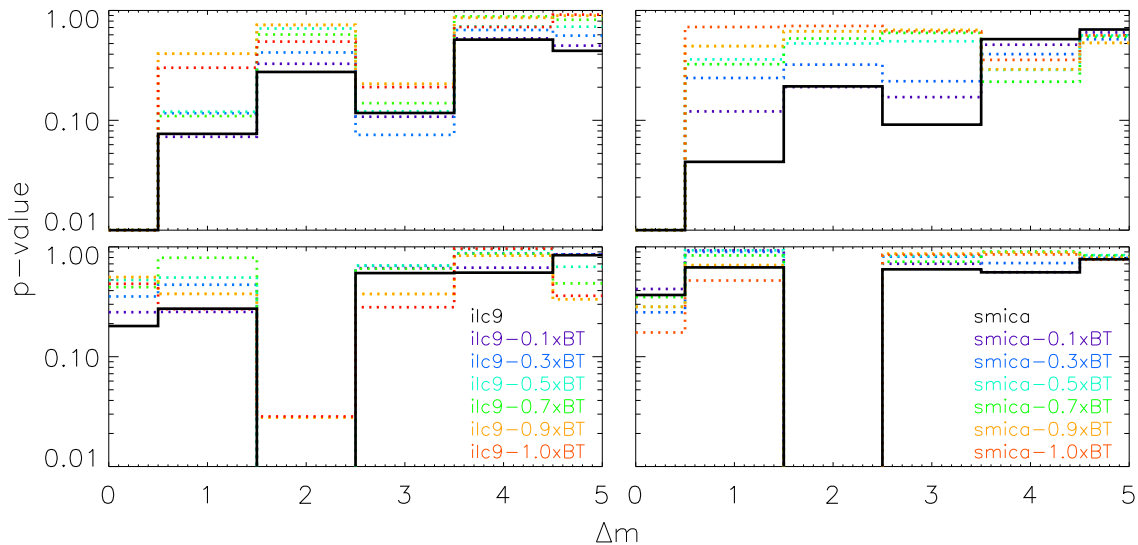}
\caption{$p$-values of the Kuiper statistic for the WMAP 9 ILC (left) and Planck SMICA map (right) (black solid line) and the corresponding Bianchi-corrected maps (colored dotted lines) for $\Delta \ell = 0$ (top) and $\Delta \ell = 1$ (bottom), calculated in the Galactic coordinate system. The Planck SEVEM map (not shown here) resembles the SMICA results. Note that subtracting the corresponding BTs does not significantly reduce the dip at $(\Delta \ell,\Delta m) = (1,2)$.}
\label{fig:kuiper}
\end{figure*}

\noindent \textbf{Method B:} Using a shuffling approach we generate surrogate maps by randomizing the potentially correlated phases $\varphi_{\ell m}$ of the original map while preserving the full-sky power spectrum of the map. In a pre-step we apply an initial Gaussian remapping of the temperature field, and a uniform remapping of the phases to avoid any influence of data outliers on the measurement of phase correlations. If the original phases are independent, the shuffling process will not influence the real-space higher order statistics of the maps. Phase correlations that are contained in the original data will be destroyed by the shuffling process. In order to enable a scale-dependent analysis of the maps we generate one first order surrogate and 500 second order surrogate maps. In the first order surrogate only the phases with $\ell$ outside the $\ell$-range of $[2,10]$ are randomized. In a second step, we shuffle the remaining phases inside that $\ell$-range. Significant deviations between these two classes of surrogates reveal phase correlations in the original data amongst $\varphi_{\ell m}$ with $\ell \in [2,10]$. In order to quantify the higher order content of the surrogate maps we use two comparable real-space image analysis methods sensitive to HOS, namely the scaling index method (SIM) developed in \cite{1997JOSAA..14.3208R,PhysRevLett.109.144101} and a set of three statistics known as the Minkowski functionals (MFs) \cite{Minkowski:1903fk}, which were introduced into cosmology by \cite{1994A&A...288..697M,1998NewA....3...75W,1998MNRAS.297..355S}. They give an $S$-value which quantifies the $\sigma$-normalized deviation between the two types of surrogates. For details and former results see \cite{2013MNRAS.428..551M}. It can be shown that method B is rotational invariant. Regardless of the chosen coordinate system, the $S$-value pattern on the sphere is preserved.

\begin{table}[h]
\centering	
\begin{tabular}{c||c|c|c|c|c|c}
& \multicolumn{2}{c|}{ILC9} & \multicolumn{2}{c|}{SMICA} & \multicolumn{2}{c}{SEVEM}\\
\hline
$(\Delta \ell, \Delta m)$ & orig & corr & orig & corr & orig & corr\\ 
\hline
(0,1) & {\bf0.075} & 0.301 &  {\bf0.042} & 0.707 & 0.160 & 0.853 \\
(0,2) & 0.275  & 0.520 & 0.204 & 0.722 &  {\bf0.096} & 0.916\\
(0,3) & 0.117 & 0.201 &  {\bf0.091} & 0.645 & 0.242 & 0.526\\
(1,2) &   {\bf0.003} &  {\bf0.029} &  {\bf0.002} &  {\bf0.005} &  {\bf0.008} &  {\bf0.060}\\
(2,5) &  {\bf0.090} & 0.193 &  {\bf0.087} & 0.471 & 0.151 & 0.454\\
(5,0) & 0.321&  {\bf0.064} & 0.148 & 0.126 & 0.132 & 0.196\\
(5,4) & 0.364 & 0.218 & 0.220 & 0.328 &  {\bf0.078} & 0.139\\
\end{tabular}
\caption{Combinations of $(\Delta \ell, \Delta m)$ with $p$-values $< 0.1$ (bold) for at least one of the three maps SMICA, SEVEM and ILC9. We show $p$ before and after the full BT correction.}
\label{tab:dldm}
\end{table}

\begin{figure*}[t!]
\includegraphics[width=15cm,keepaspectratio=true]{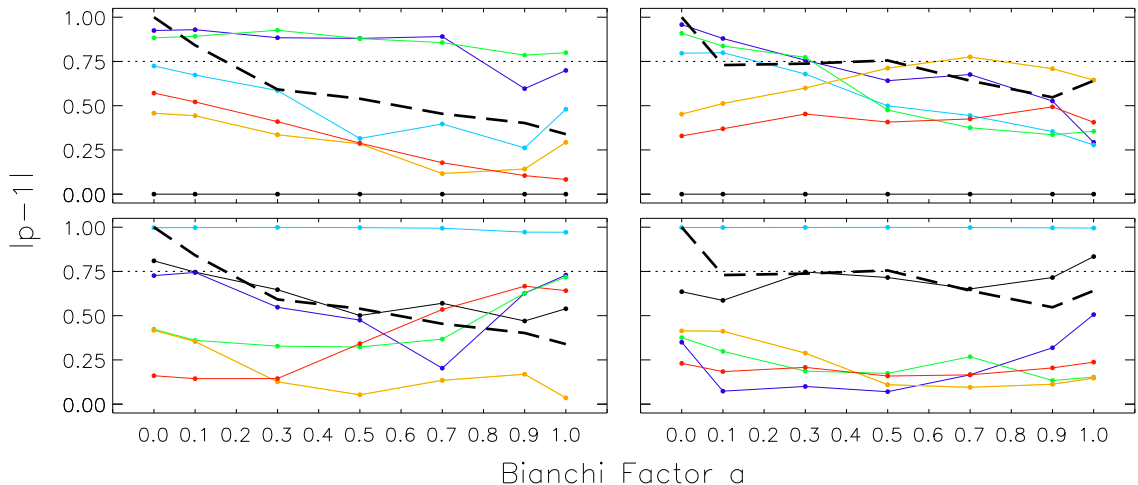}
\caption{$|p-1|$-values of the ILC9 (left) and SMICA (right) map for $\Delta \ell = 0$ (top) and $\Delta \ell = 1$ (bottom). The colored lines mark the different $\Delta m$ with black=0, purple=1, blue=2, green=3, orange=4, red=5. The black dashed line shows the normalized mean $S$-value of the output maps from method B (Figure \ref{fig:smicaMinkos}).}
\label{fig:kuiper2}
\end{figure*}

\textit{Results}.--- Comparing the results from the Kuiper statistic of 36 $(\Delta \ell, \Delta m)$-combinations with $\Delta \ell = 0-5$ and $\Delta m = 0-5$ in different coordinate systems, we find the trend that gradually subtracting the corresponding BT from the original map leads to increasing Kuiper $p$-values with increasing Bianchi factor $a$ if $p<0.1$ in the original data. This behavior is a strong indication for vanishing phase correlations due to BT correction in certain phase separation subsets depending on the chosen coordinate system. Figure \ref{fig:kuiper} shows the significance levels for $12$ combinations of $\Delta \ell = 0, 1$ and $\Delta m = 0,1,2,3,4,5$ for all steps of BT corrections with respect to the Galactic coordinate system. The majority of the $(\Delta \ell, \Delta m)$-mode-pairs has $p$-values that lie well above $0.05$ and even $0.1$ in the original map as well as after the BT corrections. These results have a relatively high probability of above $10\%$ to arise under the null-hypothesis of random phases. In rare cases, we find higher $p$-values in the original map and values below 0.1 after the BT correction. In the Galactic coordinate system e.g., this is the case at $(\Delta \ell, \Delta m) = (5,0)$ in the original ILC9 map, where the full BT correction leads to $p=0.064$. For SMICA this behavior is found for the correction with $0.3\times$BT and $0.5\times$BT, for SEVEM at correction steps $0.1, 0.3, 0.5$ and $0.7$. The phase separation $(\Delta \ell, \Delta m) = (1,2)$ is another exception. The subtraction of the BT is almost not increasing the $p$-values which remain below the $3\%$ level for SMICA and ILC9, and below $6\%$ for SEVEM, with respect to the Galactic coordinate system. In Table \ref{tab:dldm}, we list the $p$-values of the original maps and the fully BT corrected maps for all subsets with at least one value $<0.1$ either in the original map or after the full BT correction with $1.0\times$BT. The findings in the original maps are consistent with those for the WMAP 3 year data tested by \cite{2007ApJ...664....8C} and reveal an overall consistent picture of the different maps and experiments. Averaging over 768 rotated coordinate systems reveals phase correlations for $\Delta \ell=0,1$ and $\Delta m=0,1,2$ but none for larger phase separations (see supplemental material).

In Figure \ref{fig:kuiper2}, we show the inverse $|p-1|$-values with respect to the Bianchi factor $a$ for $\Delta \ell=0,1$ to visualize general tendencies in certain subsets (in Galactic system). The solid lines mark the $(\Delta \ell, \Delta m)$-combinations at different $\Delta m$.
The mode-pairs $(0,1),(0,2),(0,3)$ and $(2,5)$ (not shown) show a mostly monotonic decrease in the $p$-value with increasing $a$. The analysis of $(5,0)$ and $(5,4)$ (not shown) yields rather constant $p$-values above $0.75$. Mode-pairs with $(1,2)$ remain correlated in all maps.  

\begin{figure}
\includegraphics[width=4cm, keepaspectratio=true]{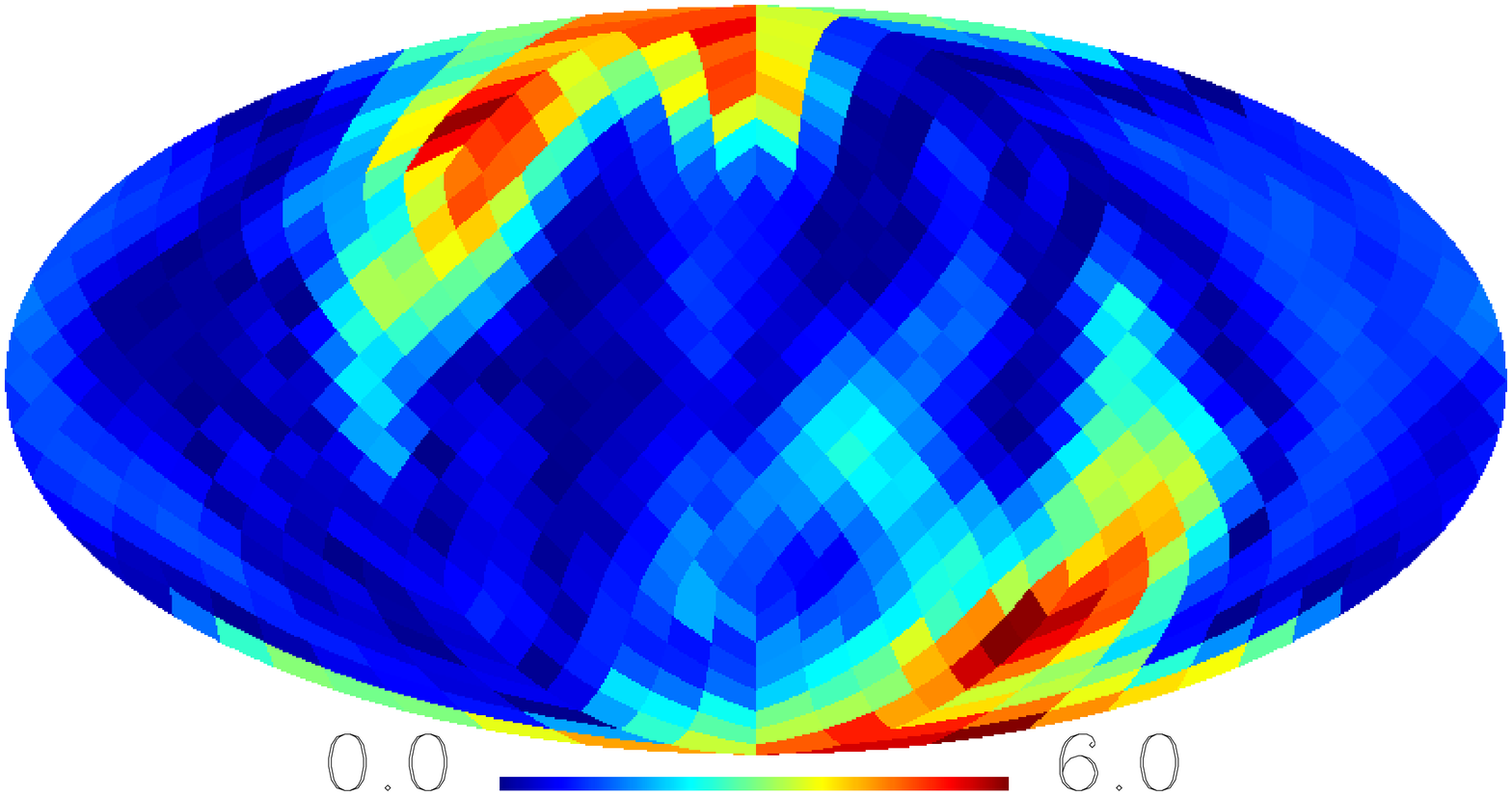}
\includegraphics[width=4cm, keepaspectratio=true]{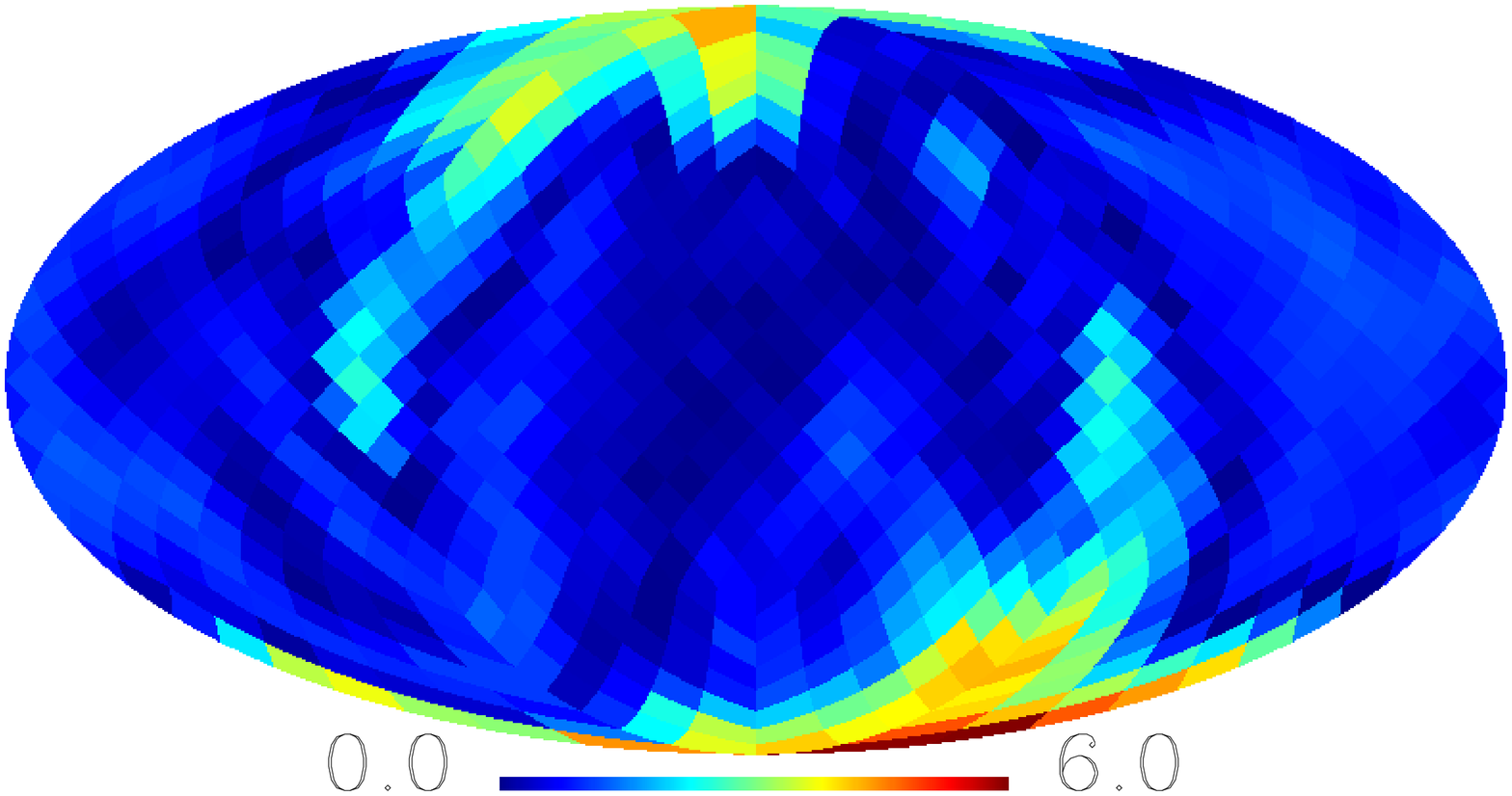}
\includegraphics[width=4cm, keepaspectratio=true]{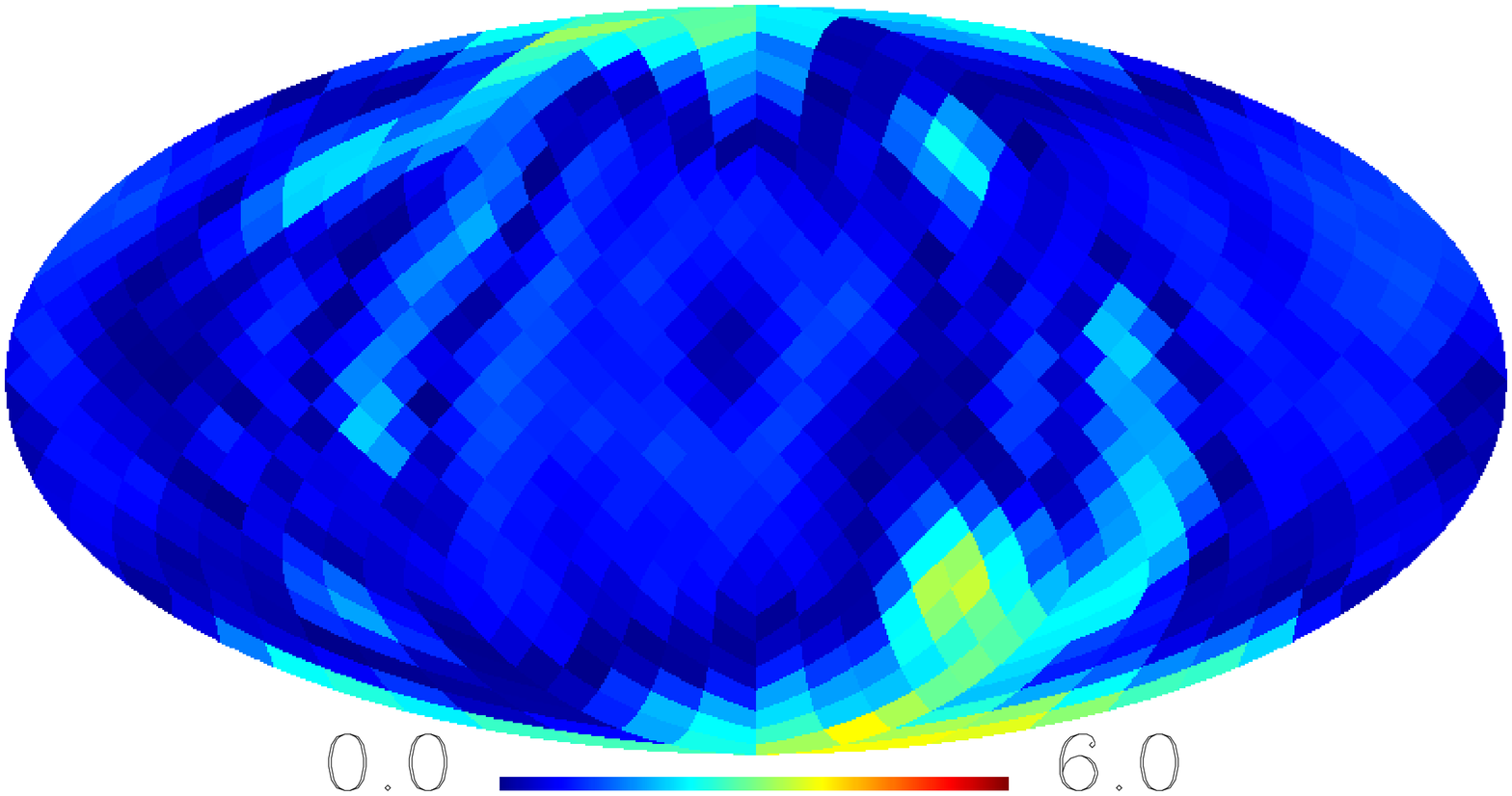}
\includegraphics[width=4cm, keepaspectratio=true]{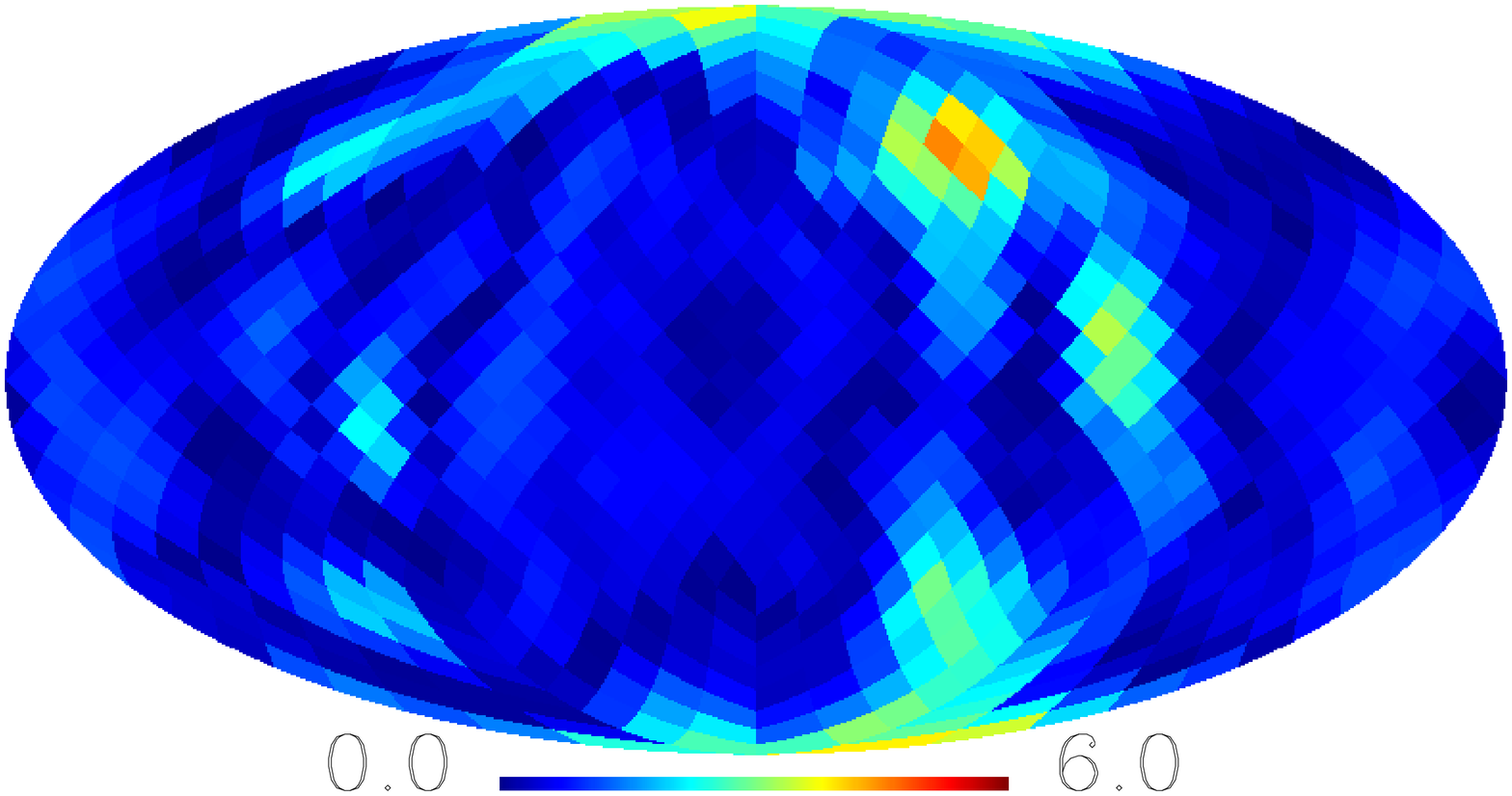}
\caption{Mollweide projection in Galactic coordinates of the $S$-value distribution of method B using the Minkowski functionals for the Planck SMICA map. The BT is subtracted with a factor of $a=0.0, 0.3, 0.7,1.0$ (upper left to lower right), respectively.}
\label{fig:smicaMinkos}
\end{figure}

Analyzing the Minkowski functionals and scaling indices from 768 different hemispheres of the sky by a $\chi^2$-statistic for surrogate 1 and 2, method B reveals that both image analysis techniques detect similar asymmetries and deviations from Gaussianity in the CMB sky. This is true for the latest release of the SMICA and SEVEM map of Planck as well as for the WMAP 9 ILC data and does not depend on the chosen coordinate system. In Figure \ref{fig:smicaMinkos}, we show the color coded $S$-values from a Minkowski analysis for the $768$ hemispheres, where a red pixel indicates strong deviations from Gaussianity in the hemisphere surrounding that pixel, and blue none. Subtracting the corresponding Bianchi type $\mathrm{VII_h}$ best-fit templates from the Planck SMICA map diminishes phase correlations and gives an increasing isotropic Gaussian sky which can be confirmed by findings for Planck SEVEM and WMAP ILC. For SMICA, the minimum signal of deviations is particularly detected at $a=0.9$, not at $1.0$. This is interesting in itself and requires further interpretation with respect to Bianchi template fitting and especially its $\ell$-range dependency. In order to quantify the overall strength of deviation from the random phase hypothesis in method B, we calculate the mean of all $S$-values and show the result in Figure \ref{fig:kuiper2}. The comparison of method A and B reveals that the normalized mean of the real-space method B decreases with $a$, likewise to the decrease for several pairs $(\Delta \ell, \Delta m) =(0,1),(0,2),(0,3),(2,5)$ in the Galactic system. Although method B reveals that the strength of the phase correlations is highly reduced after subtracting $1.0\times$BT, the $p$-value of $(\Delta \ell,\Delta m) = (1,2)$ remains almost constant. 

\textit{Conclusions \& Outlook}.---In a comparative study of the important and so far not understood relation between real-space higher order statistics and the Fourier space phase information, we provide for the first time heuristic results using the example of the spherical CMB data. Analyzing the phase distribution of CMB maps on low $\ell$-modes, we detect a clear trend, but with low statistical significance, for gradually diminished phase correlations due to the subtraction of a Bianchi type $\mathrm{VII_h}$ anisotropic cosmological template. This is especially true when looking at the subsets of ``close-by'' phase differences with $\Delta \ell=0-1$  and $\Delta m=0-3$. In comparison, we confirm a significantly vanishing higher order signal of hemispherical asymmetries in the CMB sky for Bianchi-corrected maps. We claim that the detected signatures of non-Gaussianities and hemispherical asymmetries in real-space due to phase correlations in the CMB can partly be explained by correlations between phases $\phi_{\ell m}$ separated by small $\Delta \ell$ and $\Delta m$.

Interestingly, the phase-correlations in some subsets are not diminished when subtracting the Bianchi template. Furthermore, the BT correction can induce phase-correlations for individual mode-pairs, depending on the chosen coordinate system of the map. However, the real-space signatures are not fully reduced either. The Bianchi template is not fully compatible with standard cosmological parameters and we cannot expect a perfect reduction of all existing anomalies. Moreover, the statistic of the Kuiper test with such a small effective number is not strongly significant. It is not solved yet whether individual phases $\phi_{\ell m}$ are responsible for the signatures of phase correlations or whether the relation between certain subsets plays a dominant role. On the studied very low $\ell$-range, instrumental noise is not an issue and our method is neither influenced by residual foregrounds nor experimental systematics as shown in our earlier works. We therefore expect a cosmological explanation for the detected anomalies. 
 
Our results can contribute to an understanding of non-Gaussian signals, which in cosmology may be due to Early Universe physics but might also indicate a misinterpretation of cosmological events affecting the CMB.  A detailed scale-dependent analysis and a study of the coordinate system dependencies with respect to anisotropic cosmologies can give further insight into the constraints of fully compliant cosmological models. In summary, the combined analysis of phase statistics including their variations due to either template subtraction or refined surrogate generating methods, and of the respective response of higher-order statistics offers a new statistical framework to disentangle the information content of images. 

\textit{Acknowledgements}.---We greatly thank Theresa Jaffe and Jason McEwen for providing us with the latest WMAP and Planck Bianchi templates, respectively. For the calculations we employ the HEALPix software \cite{2005ApJ...622..759G}. HIM acknowledges the support of the Christiane N\"usslein-Volhard (CNV) foundation and the International Max Planck Research School.

\bibliographystyle{apsrev4-1}
\bibliography{/Users/hmeyer/Documents/Literatur}

\end{document}


\begin{figure*}
\includegraphics[width=0.49\textwidth, keepaspectratio=true]{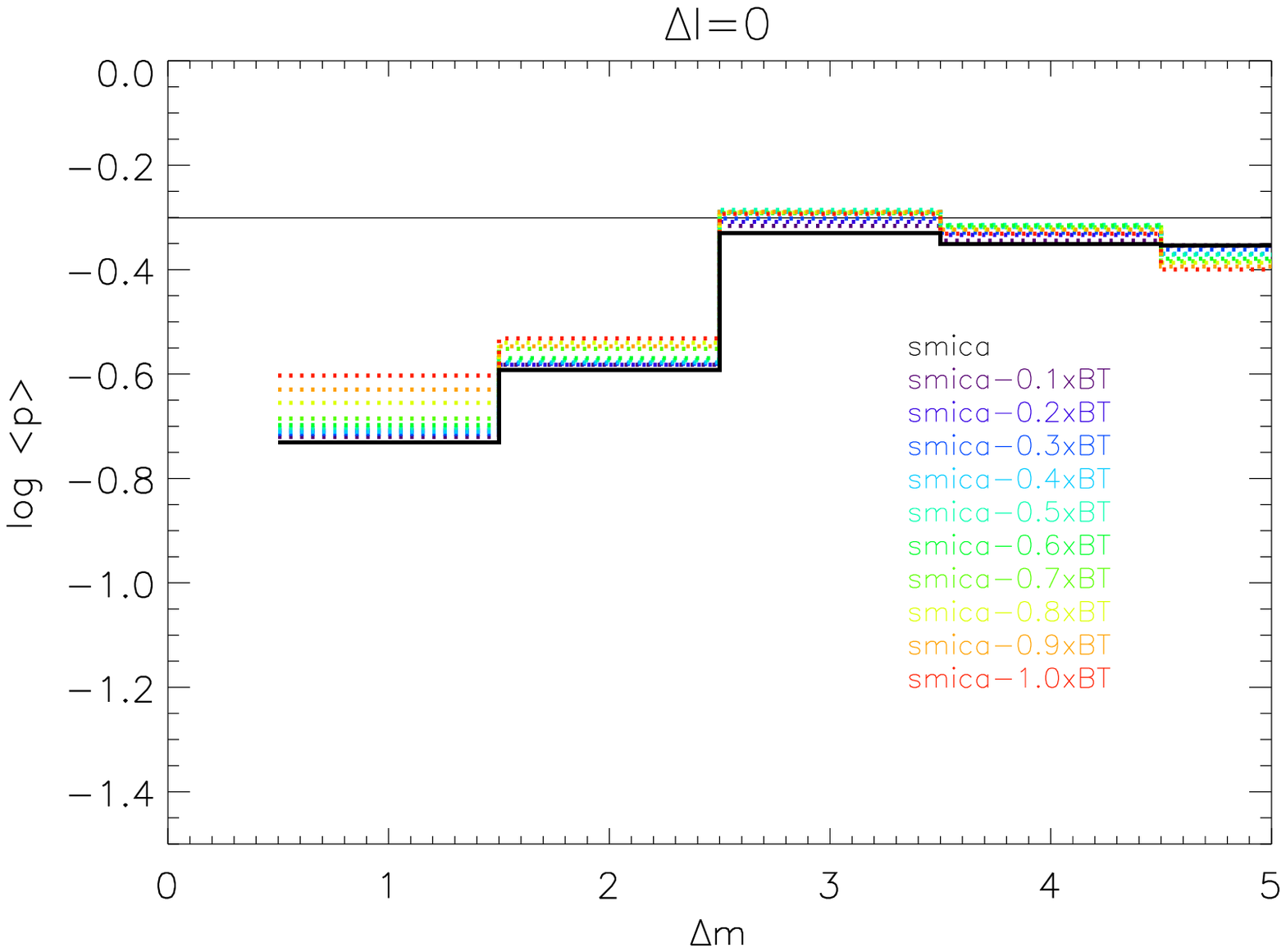}
\includegraphics[width=0.49\textwidth, keepaspectratio=true]{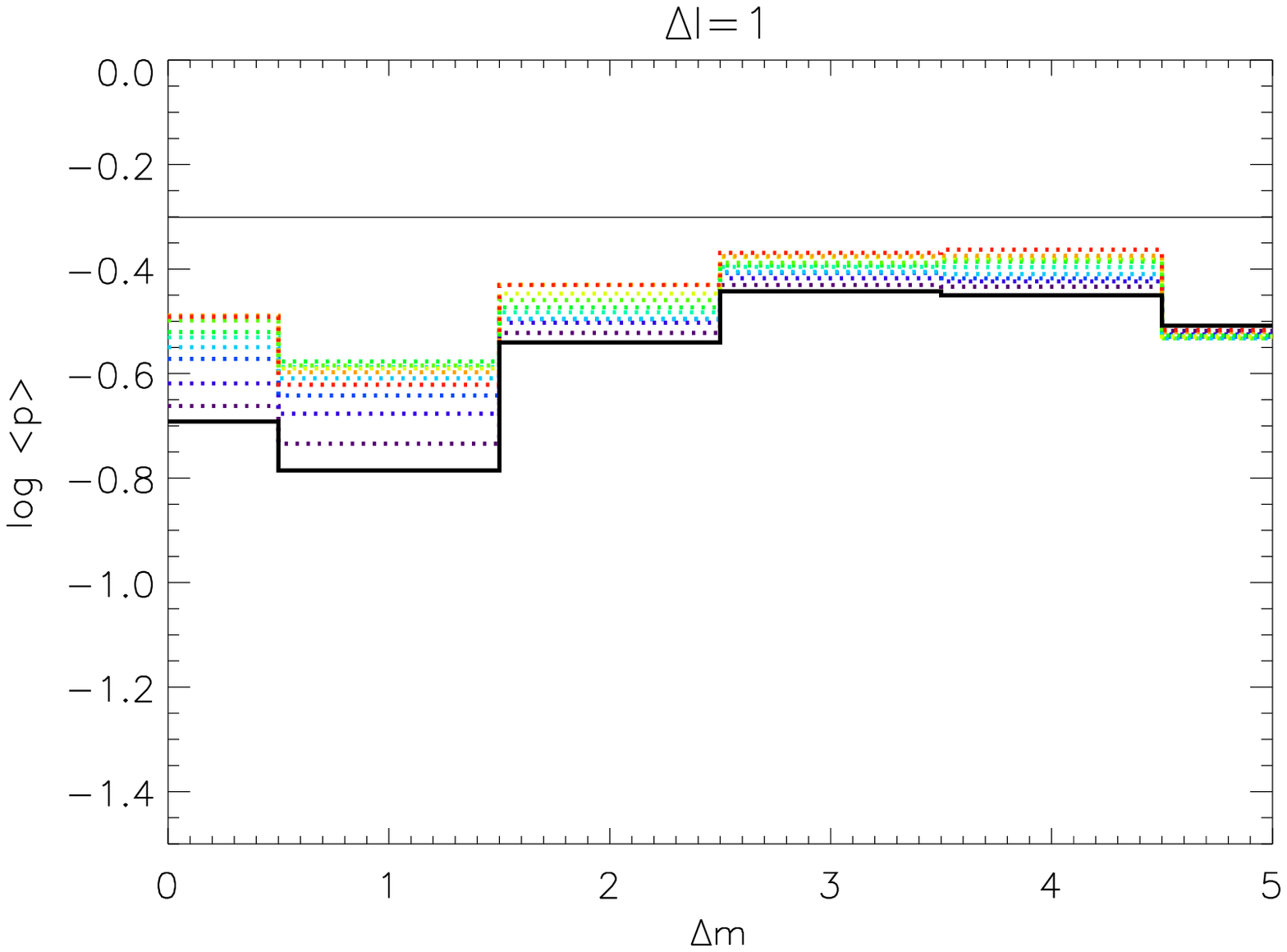}
\includegraphics[width=0.49\textwidth, keepaspectratio=true]{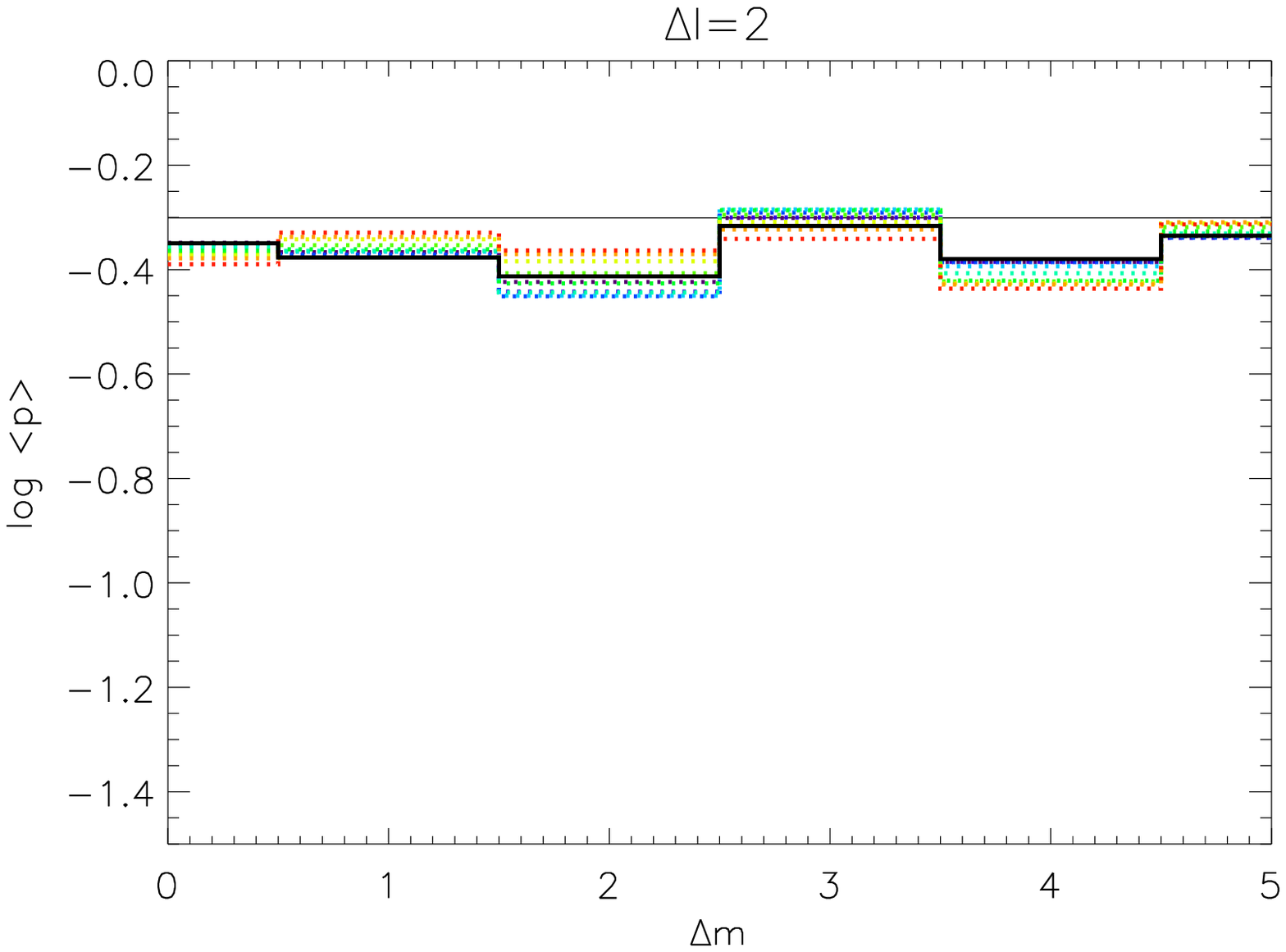}
\includegraphics[width=0.49\textwidth, keepaspectratio=true]{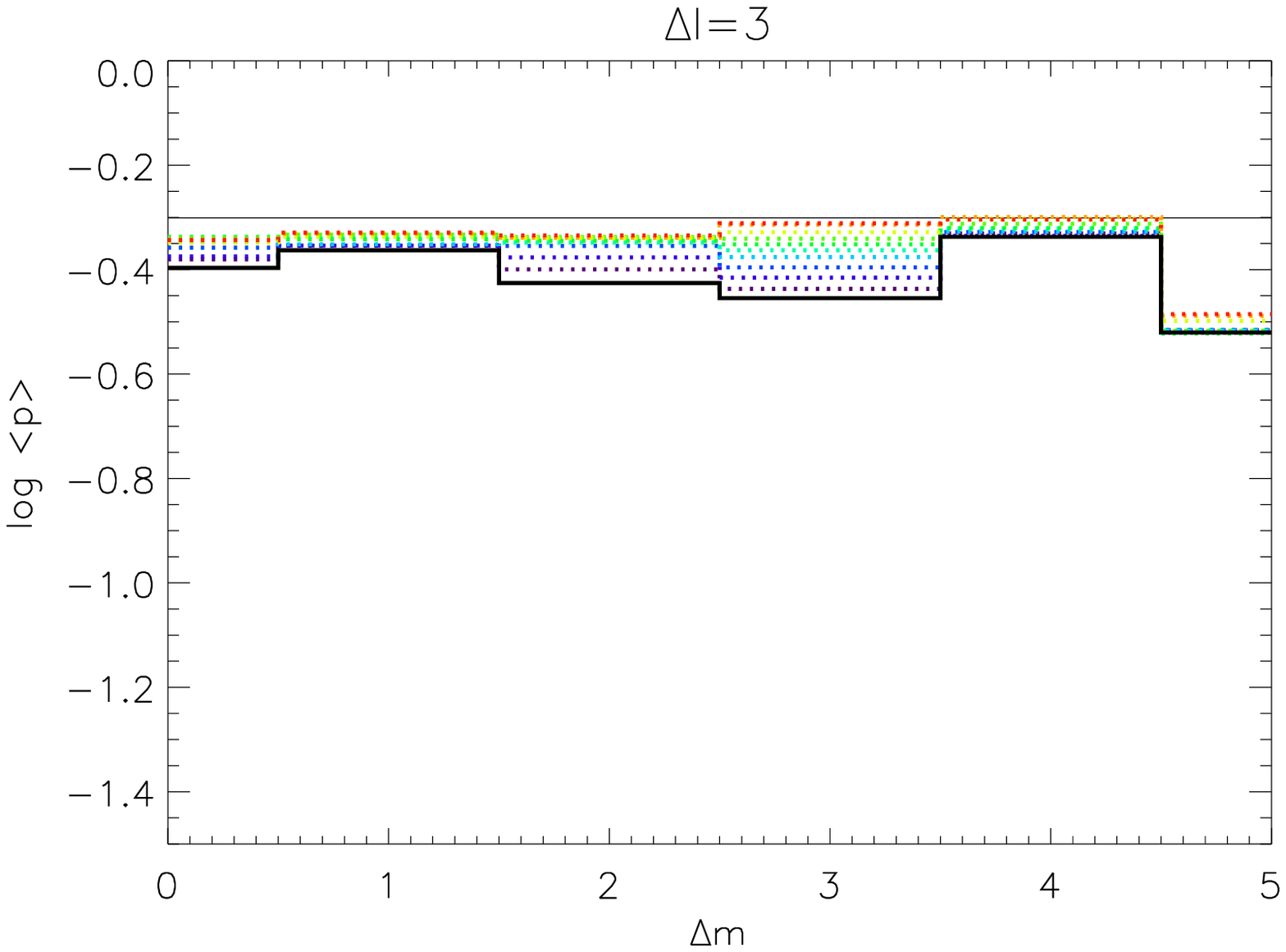}
\includegraphics[width=0.49\textwidth, keepaspectratio=true]{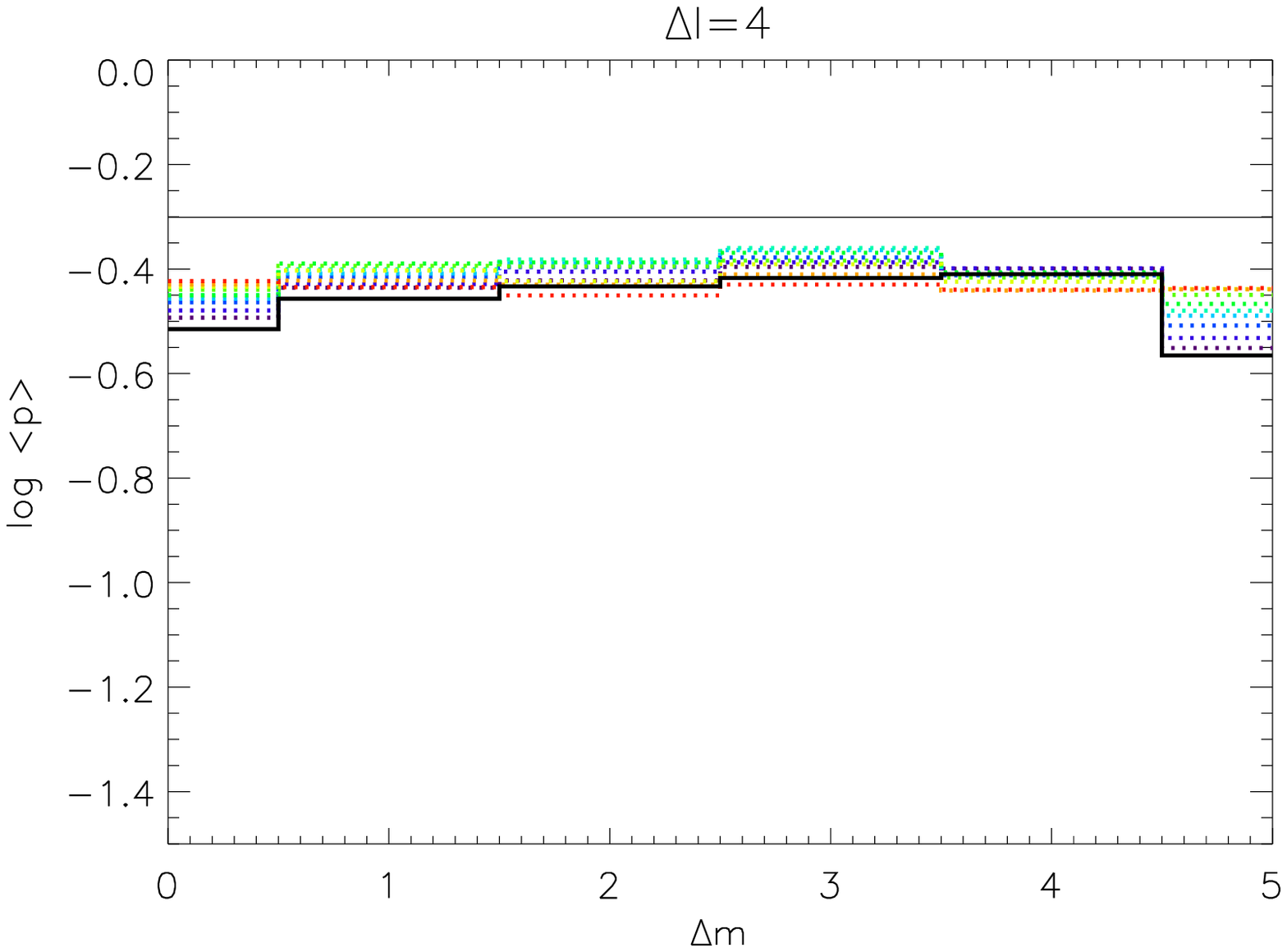}
\includegraphics[width=0.49\textwidth, keepaspectratio=true]{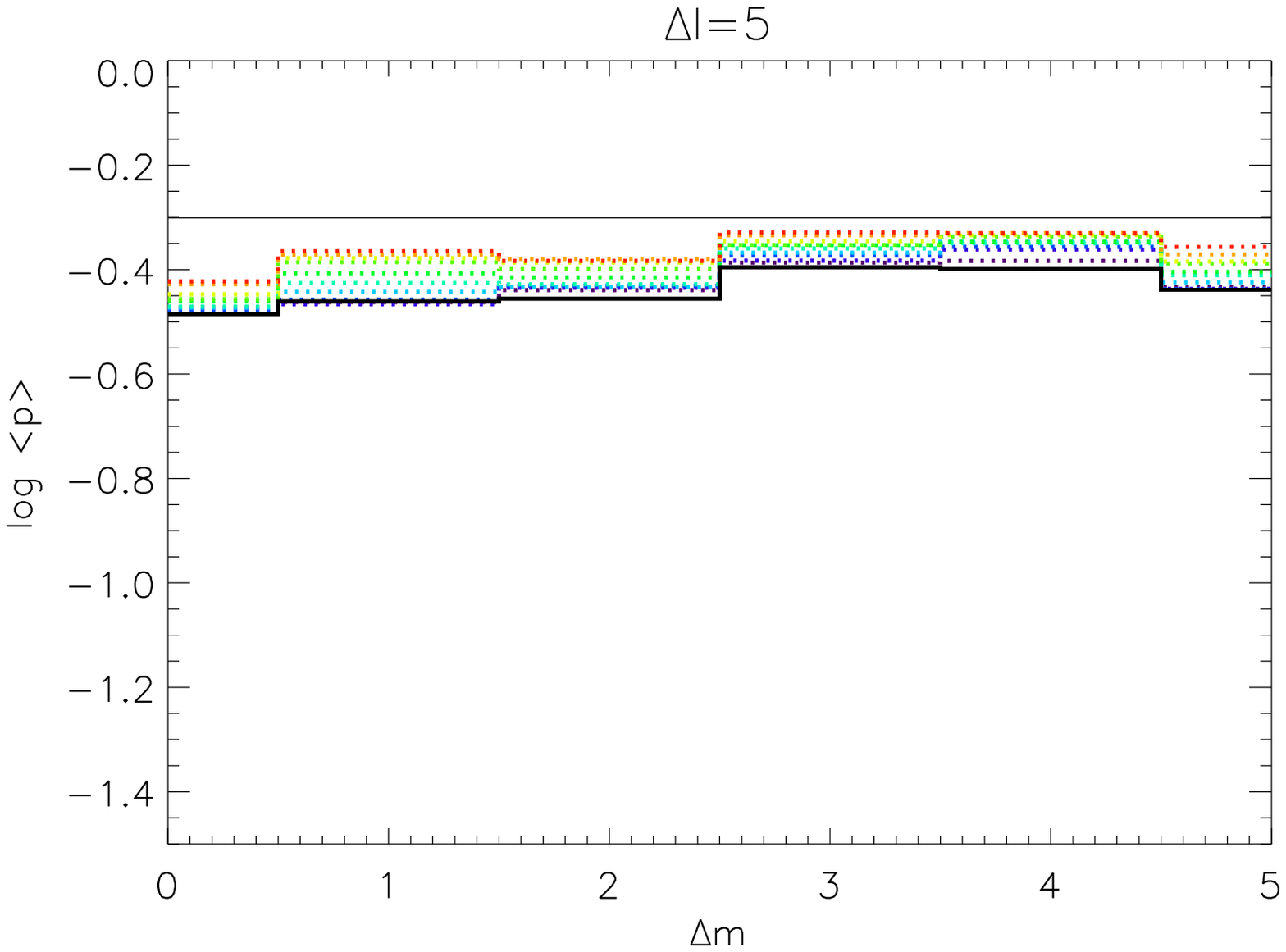}
\caption{Mean $p$-values of the Kuiper statistic for Planck SMICA map and the corresponding Bianchi-corrected maps (colored dotted lines) for $\Delta \ell=0$ to $\Delta \ell =5$ (left to right and top to bottom), averaged over 768 coordinate systems, shown in base 10 logarithm.}
\label{fig:supplementalMaterial}
\end{figure*}